\documentclass[twocolumn,amsmath,amssymb,prl,10pt,nofootinbib,superscriptaddress]{revtex4}
\usepackage{ graphicx, float,amsmath, amsmath, amssymb}
\usepackage[export]{adjustbox}

\def\be{\begin{equation}}
\def\ee{\end{equation}}
\def\bea{\begin{eqnarray}}
\def\eea{\end{eqnarray}}
\def\bse{\begin{subequations}}
\def\ese{\end{subequations}}

\usepackage[breaklinks, colorlinks, citecolor=blue]{hyperref}
\usepackage[labelsep=period]{caption}
\usepackage[normalem]{ulem}
\usepackage{soul}
\usepackage{minitoc}
\usepackage{subfigure}

\begin{document}
\title{An overview of quasinormal modes in modified and extended gravity}

\author{Flora Moulin}%
\affiliation{%
Laboratoire de Physique Subatomique et de Cosmologie, Universit\'e Grenoble-Alpes, CNRS/IN2P3\\
53, avenue des Martyrs, 38026 Grenoble cedex, France
}

\author{Aur\'elien Barrau}%
\affiliation{%
Laboratoire de Physique Subatomique et de Cosmologie, Universit\'e Grenoble-Alpes, CNRS/IN2P3\\
53, avenue des Martyrs, 38026 Grenoble cedex, France
}

\author{Killian Martineau}%
\affiliation{%
Laboratoire de Physique Subatomique et de Cosmologie, Universit\'e Grenoble-Alpes, CNRS/IN2P3\\
53, avenue des Martyrs, 38026 Grenoble cedex, France
}

\begin{abstract} 
As gravitational waves are now being nearly routinely measured with interferometers, the question of using them to probe new physics becomes increasingly legitimate. In this article, we rely on a well established framework to investigate how the complex frequencies of quasinormal modes are affected by different models. The tendencies are explicitly shown, for both the pulsation and the damping rate. The goal is, at this stage, purely qualitative. This opportunity is also taken to derive the Regge-Wheeler equation for general static and spherically symmetric metrics.
\end{abstract}
\maketitle

\section{Introduction}

General relativity (GR) is our best theory of spacetime. Although the Lovelock theorem \cite{Lovelock:1971yv} ensures that it cannot be easily modified there are quite a lot of attempts to relax some hypotheses and build a deeper model to describe the gravitational field. From effective quantum gravity to improved infrared properties, the motivations to go beyond GR are countless. So are the situations, both in astrophysics and cosmology, where extended gravity theories can, in principle be tested. In practice, reaching the level of accuracy useful to probe the relevant range of parameters is obviously far from trivial. In this article we focus on a specific aspect of gravitational waves that would be emitted during the relaxation phase of a deformed black hole (BH). \\

We will consider quasinormal modes associated with the ringdown phase of a BH merger. The modes are not strictly normal due to energy losses of the system through gravitational waves. The boundary conditions for the equation of motion are unusual as the wave has to be purely outgoing at infinity and purely ingoing at the event horizon. The time component of the radial part reads (an introductory review can be found in \cite{Chirenti:2017mwe})

\begin{equation}
e^{-i\omega t} = e^{-i(\omega_R + i\omega_I)t},
\end{equation}
 the complex pulsation $\omega$ being split in a real part $\omega_R$, which corresponds to the frequency, and an imaginary one $\omega_I$, which is the inverse timescale of the damping. Stability requires $\omega_I<0$. Although real-life BHs are spinning, we focus on Schwarzschild solutions in this article. The details of these predictions can not be used to directly compare with observations. We, however, expect the general tendencies and orders of magnitudes to remain correct, as it can be checked for the general relativistic case in \cite{Krivan:1997hc}.\\

The linearized Einstein equations lead to wave equations with different potentials whether one considers ``axial" or ``polar" perturbations. In GR, the (so called Regge-Wheeler) potential for axial perturbations is
\begin{equation}
V^{\textrm{RG}}_{\ell}(r) = \left(1-\frac{2M}{r}\right)\left[\frac{\ell(\ell+1)}{r^2} - \frac{6M}{r^3}\right]\,,
\label{eq:RG}
\end{equation}
while the (so-called Zerilli) one for polar perturbations is
\begin{eqnarray}
& &V^{\textrm{Z}}_{\ell}(r) = \frac{2}{r^3}\left(1-\frac{2M}{r}\right)\times \nonumber\\
&\times&\frac{9M^3 + 3a^2Mr^2 + a^2(1+a)r^3 + 9M^2ar}{(3M+ar)^2}\,,
\label{eq:Ze}
\end{eqnarray}
where $a = \ell(\ell+1)/2 - 1$. Throughout all the paper we use Planck units. In the purely gravitational sector, one needs $\ell\ge2$. Interestingly, both those equations have the very same spectrum of quasinormal modes (QNMs). This property, called isospectrality \cite{Chandrasekhar:1985kt} is not always true in modified gravity (see \cite{Moulin:2019bfh} for an extension and a discussion of the original proof). Basically, quasinomal modes are described by their multipole number $\ell$ and their overtone number $n$. The fundamental quadrupolar mode ($n=0$ and $\ell=2$) for a Schwarzschild BH in GR is given by $M\omega \approx 0.374 - 0.0890i$. \\

There are many different ways to calculate the QNMs: continued fractions, Frobenius series, Mashhoon's method, confluent Heun’s equation, characteristic integration, shooting, WKB approximations, etc. In this article we focus on the last approach. For most models considered here, the QNMs have already been calculated in previous studies. However, this has most of the time been done for $s=0$ or $s=1$, not for $s=2$ as we have done it here. More importantly, it is in addition very useful to rely on the very same method to investigate all models so that the differences underlined are actually due to physical effects and not to numerical issues. Even when the same approach is considered, the way it is implemented is often different enough, between articles, so that it is hard to directly compare the results. This is why we have here tried to consider methodically several modified gravity models with a well controlled WKB approximation scheme used in the same way in all cases so as to compare the tendencies between modified gravity proposals. This is not mandatory for this qualitative step but this will become useful in future quantitative studies.\\

The determination of the complex frequencies of QNMs is difficult (see \cite{Kokkotas:1999bd,Nollert:1999ji} for historical reviews and \cite{Berti:2004um,Dorband:2006gg} for results based on numerical approaches). This work is based on the WKB approach described in \cite{Konoplya:2003ii}. Following the pioneering work \cite{Mashhoon:1982im}, the WKB method for QNMs was developed in \cite{Schutz:1985zz,Iyer:1986np,Iyer:1986nq,Kokkotas:1988fm}. This formalism leads to fairly good approximations, especially for high multipole and low overtone numbers. In the following, we restrict ourselves to $n<l$ and use the 6th order WKB method developed by Konoplya \cite{Konoplya:2003ii} (see also \cite{Konoplya:2009ig,Konoplya:2011qq,Konoplya:2019hlu}). This allows one to recast the potential appearing in the effective Schr\"odinger equation felt by gravitational perturbations in a complexe but tractable form. \\

The aim of this introductory paper is to investigate how several modified gravity theories impact the QNMs at the qualitative level. There are several ways to go beyond GR: extra dimensions, weak equivalence principe violations, extra fields, diffeomorphism-invariance violations, etc. Beyond those technicalities, there are strong conceptual motivations to consider extended gravity approaches, from the building of an effective quantum gravity theory to the improvements of the renormalisation properties, through the implementation of a dynamical cosmological constant. Among many others, examples of recent relevant works on QNMs can be found in \cite{Blazquez-Salcedo:2016enn,Blazquez-Salcedo:2017txk,Chen:2018vuw,Chen:2019iuo}.

\section{Perturbation dynamics}

The QNMs are solutions of a perturbation equation with the specific boundary conditions given in the previous section. The radial and angular parts can be separated. The radial part is governed by a Schr\"odinger-like equation:

\be
\frac{d^2 Z}{dr^{*2}}+ V(r) Z=0,
\label{schrolike}
\ee

\noindent  where Z is the radial part of the ``perturbation" variable, assumed to have a time-dependance $e^{i \omega t}$, and $r^*$ is the tortoise coordinate. For a metric such that

\be
ds^2=f(r)dt^2-f(r)^{-1}dr^2-r^2d \theta ^2 -r^2 \sin ^2 \theta
d \phi ^2, 
\label{metric1}
\ee

\noindent the tortoise coordinate is defined by

\be 
dr^*=\frac{1}{f(r)}dr.
\label{tortoise1}
\ee
It tends to $- \infty $ at the event horizon and to $+ \infty$ at spatial infinity.\\

As explained previously, BH gravitational perturbations can be of two different types distinguished by their behavior under a parity transformation. For an angular momentum $l$, axial perturbations transform as $(-1)^l$ under parity, while polar perturbations transform as $(-1)^{l+1}$. This leads to the two different potentials in Eq.(\ref{schrolike}). The potentiel for the gravitational axial perturbations reads in full generality (see \cite{Chirenti:2017mwe} and references therein) for the metric given by Eq. (\ref{metric1}):

\be 
V(r)=f(r) \left(\frac{\lambda + 2 (f(r) -1)}{r^2}-\frac{f'(r)}{r} \right).
\label{pot}
\ee

In this work we will not consider the isospectrality-violation issues and we will focus only on such perturbations. It should anyway be kept in mind that, in principle, isospectrality might not hold. \\

The boundary conditions can be expressed as

\begin{eqnarray}
Z \sim e^{-i \omega r^*} \ \ \ \ \ \ r^* \rightarrow - \infty, \\
Z \sim e^{i \omega r^*} \ \ \ \ \ \ r^* \rightarrow + \infty.
\label{BC}
\end{eqnarray}

We shall now derive the Regge-Wheeler equation for the more general (spherical and static) metric:

\be
ds^2=A(r)dt^2-B(r)^{-1}dr^2-H(r)d \theta ^2 - H(r) \sin ^2 \theta
d \phi ^2 . 
\label{metric}
\ee

For this metric, the tortoise coordinate is defined by

\be 
\frac{d}{dr^*}= \sqrt{AB}\frac{d}{dr}. 
\label{tortoise2}
\ee

The general form of an axisymmetric metric can be written as \cite{Chandrasekhar:1985kt}: 

\bea
ds^2=&e^{2 \nu } (dx^0)^2-e^{2 \psi } (d x^1  - \sigma dx^0 - q_2 dx^2 - q_3 d x^3)^2 \nonumber \\
&- e^{2 \mu _2 } (dx^2)^2-e^{2 \mu _3 }(dx^3) ^2 ,
\label{generalmetric}
\eea

\noindent where $t=x^0$, $\phi = x^1$, $r=x^2$ and $\theta = x^3$. For the metric given by Eq. (\ref{metric}), the correspondance is:

\bea
&& e^{2 \nu } =A(r),  \quad 
e^{-2 \mu _2 }=B(r),  \nonumber  \\
&& e^{2 \mu _3 }= H(r),  \quad
e^{2 \psi } = H(r) \sin ^2 \theta , \\
&& \sigma = q_2= q_3= 0. \nonumber
\eea

A perturbation of this kind of spacetime is described by $\sigma$, $q_2$ and $q_3$, assumed to be first order quantities, and by infinitesimal increments,  $\delta \nu$, $\delta  \mu _2$, $\delta  \mu _3$, of the other quantities. We focus here on axial perturbations. The point is to linearize the field equations about the solution given by Eq. (\ref{metric}), considering components where  $\sigma$, $q_2$ and $q_3$ are only function of $t$, $x^2$ and $x^3$. The equations governing $\sigma$, $q_2$ and $q_3$ are described by the vanishing of the Ricci tensor components:

\be
R_{12}=R_{13}=0.
\ee

For Eq. (\ref{generalmetric}), one has \cite{Chandrasekhar:1985kt} 
\bea
R_{12}=\frac{1}{2}e^{- 2\psi - \nu - \mu _3 } \times \nonumber \\ 	\lbrack (e^{ 3 \psi - \nu - \mu _2 + \mu _3 }Q_{02})_{,0} - (e^{3 \psi + \nu - \mu _2 - \mu _3 }Q_{32})_{,3}	\rbrack,
\label{R12}
\eea

with 
\be
Q_{ab}=q_{a,b}-q_{b,a} \quad  \mathrm{and}  \quad Q_{a0}= q_{a,0}-\sigma_{,a} \quad  \mathrm{for}  \quad a,b=2,3.
\ee

\noindent The comma indicates the derivative. The notation $Q_{0a}$ is used to mean $-Q_{a0}$.
The component $R_{13}$ is also given by Eq. (\ref{R12}) by switching indices $2$ and $3$. 

The perturbed field equation are obtain by $\delta R_{\alpha \beta} =0 $. After replacing $\nu$, $\mu _2$, $\mu _3$ and $\psi$ by their expressions,  $\delta R_{12} =0 $ leads to

\be 
(H\sin^3 \theta \sqrt{AB}
Q_{23}),_3= -H^2 \sin^3 \theta \sqrt{\frac{B}{A}} Q_{02,0}.
\ee

\noindent  By defining

\be 
Q=\sqrt{AB}HQ_{23} \sin^3 \theta,
\ee

\noindent  one obtains 

\be 
\sqrt{\frac{A}{B}}\frac{1}{H^2 \sin^3 \theta } \frac{\partial Q}{ \partial \theta}= Q_{20,0}.
\label{dqdtheta}
\ee

\noindent  For $\delta R_{13} =0 $, one is led to

\be 
\frac{\sqrt{AB}}{H \sin^3 \theta} \frac{\partial Q}{ \partial r}= - Q_{30,0}.
\label{dqdr}
\ee

\noindent  We assume that perturbation have a time dependance given by 
$
e^{i \omega t}. 
$
This implies that Eqs. (\ref{dqdtheta}) and (\ref{dqdr}) read

\be 
\sqrt{\frac{A}{B}}\frac{1}{H^2 \sin^3 \theta } \frac{\partial Q}{ \partial \theta}= - \omega^2 q_2 -i \omega \sigma _{,2},
\label{one}
\ee

\be 
\frac{\sqrt{AB}}{H \sin^3 \theta} \frac{\partial Q}{ \partial r}=  \omega ^2 q_3 +i \omega \sigma _{,3}.
\label{two}
\ee

Taking the derivative of Eq.(\ref{one}) with respect to $\theta$, the derivative of Eq. (\ref{two}) with respect to $r$, and combining the results leads to:

\be
 \sin^3 \theta \frac{\partial }{ \partial \theta}\bigg( \frac{1}{ \sin^3 \theta}\frac{\partial Q}{ \partial  \theta} \bigg) + H^2 \sqrt{\frac{  B}{A}}\frac{\partial }{ \partial r} \bigg( \frac{\sqrt{AB}}{H } \frac{\partial Q}{ \partial r}\bigg) + \omega ^2 \frac{Q  H}{A}=0.
 \label{eq}
\ee

As suggested in \cite{Chandrasekhar:1985kt}, one can then separate the variables $r$ and $\theta$ using

\be 
Q(r, \theta) = R(r) C^{-3/2}_{l+2} (\theta) 
\label{qrc}
\ee

\noindent with $C^m_n$ the Gegenbauer function satisfying

\be 
\bigg( \frac{d}{d \theta} \sin^{2m} \theta \frac{d}{d \theta } + n(n+2m) \sin^{2m} \theta \bigg) C^m_n ( \theta ) = 0.
\ee

Inserting Eq. (\ref{qrc}) into Eq. (\ref{eq}), one is led to following radial equation: 

\be 
H^2 \sqrt{\frac{  B}{A}} \frac{\partial }{ \partial r} \bigg( \frac{\sqrt{AB}}{H } \frac{\partial R(r)}{\partial r} \bigg) + \bigg( \frac{  H}{A} \omega ^2 - \mu ^2 \bigg) R(r)= 0,
\ee

\noindent where $\mu ^2= (l-1)(l+2)$. Defining $Z$ so that $R=\sqrt{H}Z$ and  using the tortoise coordinate, we are led  to a Schr\"odinger-like equation:

\be 
\frac{d^2 Z}{dr^{*2}} + (\omega ^2 - V(r) )Z =0, 
\ee

\noindent where the potential is 

\be 
V(r)= \frac{1}{2 H^2} \bigg( \frac{d H}{dr^*} \bigg) ^2 + \frac{\mu ^2 A}{H} - \frac{1}{\sqrt{H}} \frac{d^2}{dr^{*2}} \bigg( \sqrt{H} \bigg).
\label{RGHG}
\ee

The potential reduces to Eq. (\ref{pot}) for $A(r)=B(r)$ and $H(r)=r^2$. This derivation is useful to calculate QNMs for general static and spherically symmetric metrics.

\section{The WKB approximation}

The WKB approximation \cite{Schutz:1985zz,Iyer:1986np,Iyer:1986nq} is known for leading to good approximations (compared to numerical results) for the QNMs. The potential is written using the tortoise coordinate so as to be constant at $r^* \rightarrow 0$ (which represent the horizon of the BH) and at $ r^* \rightarrow + \infty $ (which represents spatial infinity). The maximum of the potential is reached at $r_0^*$. Three regions can be identified: region $I$ from $ - \infty$ to $r_1$, the first turning point (where the potentiel vanishes), region $II$ from $r_1$ to $r_2$, the second turning point, and region $III$ from $r_2$ to $+\infty$. In region $II$, a Taylor expansions is performed around $r_0^*$. In regions $I$ and $III$, the solution is approximated by an exponential function:

\be
Z \sim exp \bigg[ \frac{1}{\epsilon}\sum _{n=0} ^{\infty} \epsilon ^n S_n(x) \bigg],  \quad \quad \epsilon \rightarrow 0.
\ee

This expression can be inserted into Eq. (\ref{schrolike}) so as to obtain $S_j$ as a function of the potential and its derivative. We then impose the boundary conditions given by Eq. (\ref{BC}) and match the solutions of regions $I$ and $III$ with the solution for region $II$ at the turning points $r_1$ and $r_2$ (respectively). The WKB approximation has been usefully extended from the third to sixth order in \citep{Konoplya:2003ii}. \\

This allows one to derive  the complex frequencies as a function of the potential and its derivatives evaluated at the maximum. For the sixth order treatment, one is led to:

\be
\omega ^2= V_0 -i \sqrt{-2V_0''} \bigg( \sum _{j=2}^6 \Lambda _j + n +\frac{1}{2} \bigg),
\ee 

\noindent where the expressions of the $\Lambda _j$s can be found in \citep{Konoplya:2003ii}. In the following, we use this scheme to compare different modified gravity models and we present results only in the range of validity of the WKB approximations.\\

Interesting recent considerations on the convergence on the WKB series are given in \cite{Hatsuda:2019eoj}. Details on the expansion parameter used in this work can be found in \cite{Konoplya:2011qq}. The consistency of the WKB approximation has been checked for the presented results.

 \section{Modified gravity models and results}
 
 Throughout all this section we investigate some properties of the QNMs for several extended gravity approches. We pretend, in no way, to do justice to the subtleties of those models and, when necessary, we explicitly choose a specific of simplified setting to make the calculations easily tractable. \\
 
As we focus on phenomenological aspects, the more interesting mode is the fundamental one: $n=0$ and $l=2$. We therefore focus on a few points around this one (keeping in mind that the accuracy is better for higher values of $l$). In all the figures, the lower overtone $n$ is the one with the smallest imaginary part. \\

 We first consider models with a metric of the form:
 
 \be
ds^2=f(r)dt^2-f(r)^{-1}dr^2-r^2d \theta ^2 -r^2 \sin ^2 \theta
d^2 \phi , 
\ee

\noindent and, then, investigate a model with two different metric functions, using the result obtained in Eq. (\ref{RGHG}).
 
 \subsection{1. Massive gravity}
 
 In GR, the graviton is a massless spin-2 particle. One of the first motivations for modern massive gravity -- which can be seen as a generalization of GR -- was the hope to account for the accelerated expansion of the Universe by generating a kind of Yukawa-like potential for gravitation \cite{DAmico:2011eto}. The initial linear approach to massive gravity was containing a Boulware-Deser ghost, which was cured in the dRGT version \cite{deRham:2010ik,deRham:2010kj,Hassan:2011vm,deRham:2014zqa}. Massive gravity also features interesting propertied for holography (see, {\it e.g.} \cite{dS:2010fh}). 

Starting from the action
\begin{equation}
S=\frac{1}{16\pi }\int \text{d}^{4}x\sqrt{-g}\left( R+m^{2}\mathcal{U}(
g,\phi ^{a}) \right) ,  \label{action}
\end{equation}%
where $R$ is the Ricci scalar and $\mathcal{U}$ is the potential for the graviton, the following black hole solution can be derived \cite{Ghosh:2015cva,EslamPanah:2018rob}:
  
 \be 
 f(r)= 1-\frac{2M}{r}+\frac{\Lambda r^2}{3}+ \gamma r + \epsilon , 
 \ee

\noindent where $\Lambda$, $\gamma$ and $\epsilon$ are, respectively

\bea
\Lambda = 3m^2(1+a + b ), \nonumber \\
\gamma = -cm^2(1+2 a + 3b ), \nonumber \\
\epsilon = c^2m^2(a+ 3b ),
\eea

\noindent $a$ and $b$ being two  dimensionless constants and $c$ is positive. It should also be pointed out that a positive value of $\gamma$ might raise consistency issues \cite{EslamPanah:2018rob}.\\

  \begin{figure}
\centering
    \includegraphics[width=.9\linewidth]{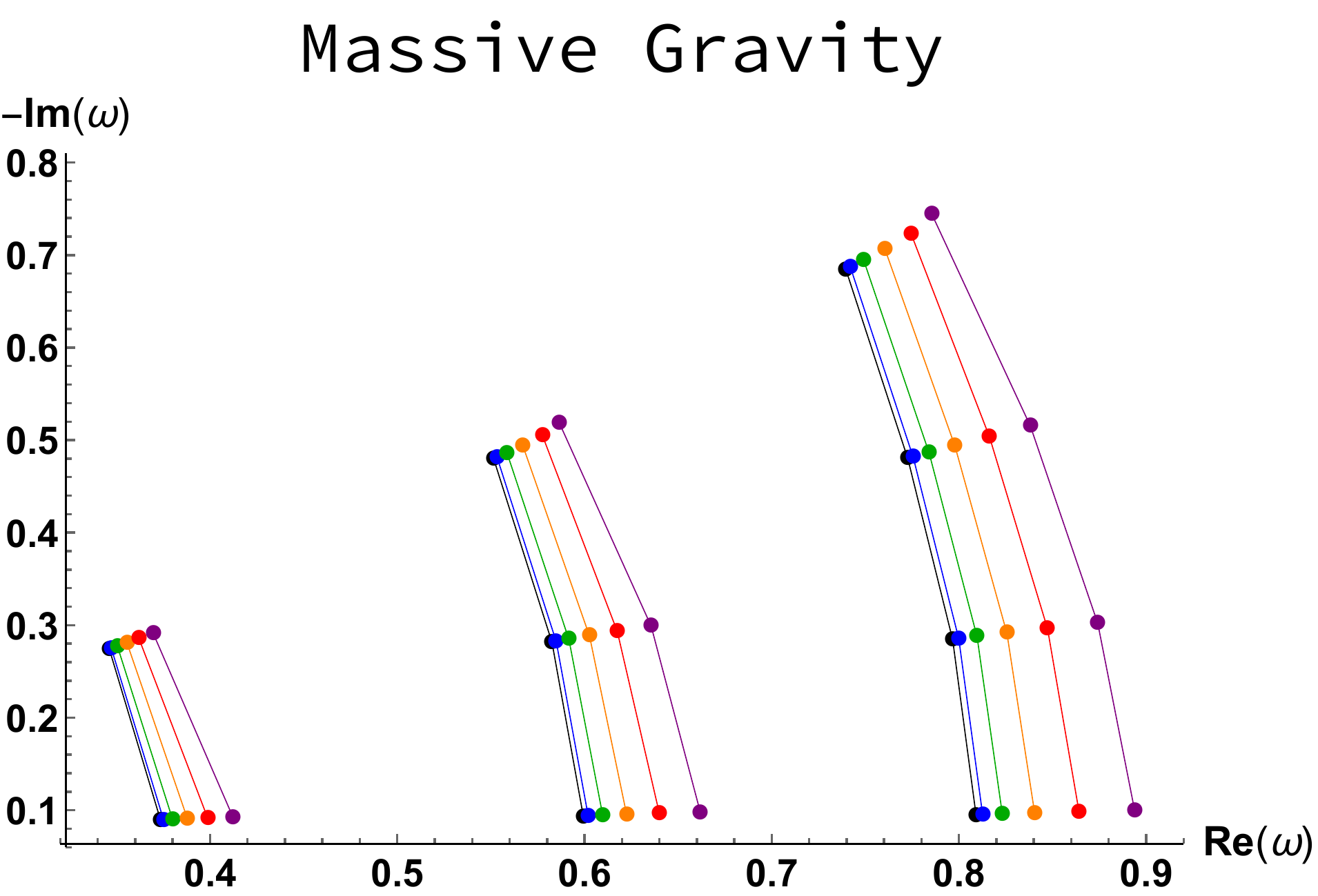}  
 \caption{QNMs in massive gravity. The left block is for $l=2$, the middle one corresponds to $l=3$ and the right one is for $l=4$. The dark points correspond to the Schwarzschild QNMs. The arbitrary constants $a$, $b$ and $c$ have been taken to one. From left to right: $m= \{15, 30, 45, 60, 75 \} \times 10^{-3}$.  }
 \label{Mass}
  \end{figure}

The results are presented in Fig. \ref{Mass}. The values chosen for the constants do of course change the amplitude of the displacement of the QNMs. The global trend, which is the point of this study, however remains the same. Increasing of the graviton mass $m$ tends to increase the real part of QNMs, that is the frequency of the oscillations. The difference in frequency between the fundamental and the first overtone also increases with $m$. The effect on the imaginary part is hardly noticeable on the plot even though a slight increase should be noticed, which is actually 50\% less important, in relative variation, than the shift in frequency. The values considered here for the mass are, of course, way out of the known bounds but this is clearly not the point. As a specific feature, one can notice that the frequency shift due to massive corrections decreases for higher overtones. The shift patterns are mostly the same whatever the multipole number considered. 

\subsection{2. Modified STV gravity}

The Scalar-Tensor-Vector modified gravitational theory (MOG) allows the gravitational constant, a vector field coupling, and the vector field mass to vary with space and time \cite{Moffat:2005si}. The equations of motion lead to an effective modified acceleration law that can account for galaxy rotation curves and cluster observation without dark matter. Although it has recently been much debated and put under pressure, the theory is still worth being considered seriously. We consider the field equation for the metric tensor \cite{Moffat:2014aja} :
\begin{equation}
R_{\mu\nu}=-8\pi GT_{\phi\mu\nu},
\end{equation}
where the gravitational coupling is $G=G_N(1+\alpha)$, with $G_N$ the Newton's constant. The gravitational strength of the vector field $\phi_\mu$ (spin 1 graviton) is $Q_g=\sqrt{\alpha G_N}M$.
With $B_{\mu\nu}=\partial_\mu\phi_\nu-\partial_\nu\phi_\mu$, the energy-momentum tensor for the vector field is :
\begin{equation}
T_{\phi\mu\nu}=-\frac{1}{4\pi}({B_\mu}^\alpha B_{\nu\alpha}-\frac{1}{4}g_{\mu\nu}B^{\alpha\beta}B_{\alpha\beta}),
\end{equation}
the constant $\omega$ of  \cite{Moffat:2005si} being set to one. Solving the vacuum field equations 

\begin{equation}
\label{Bequation}
\nabla_\nu B^{\mu\nu}=\frac{1}{\sqrt{-g}}\partial_\nu(\sqrt{-g}B^{\mu\nu})=0,
\end{equation}
and
\begin{equation}
\label{Bcurleq}
\nabla_\sigma B_{\mu\nu}+\nabla_\mu B_{\nu\sigma}+\nabla_\nu B_{\sigma\mu}=0,
\end{equation}

\noindent with the appropriate symmetry leads to the metric 

 \be 
 f(r)= 1-\frac{2M}{r}+\frac{\alpha ( 1 + \alpha)M^2}{r^2}. 
 \ee

We focus on the case where the field equations for $B_{\mu\nu}$ are non-linear, as the phenomenology is then richer, and we consider the relevant choice $\alpha < \alpha _c= 0.67$ where there are two horizons and an appropriate potential behavior for the WKB approximation to hold. An up-to-date investigation of QNMs in MOG can be found in \cite{Manfredi:2017xcv}.\\

\begin{figure}
\centering
    \includegraphics[width=.9\linewidth]{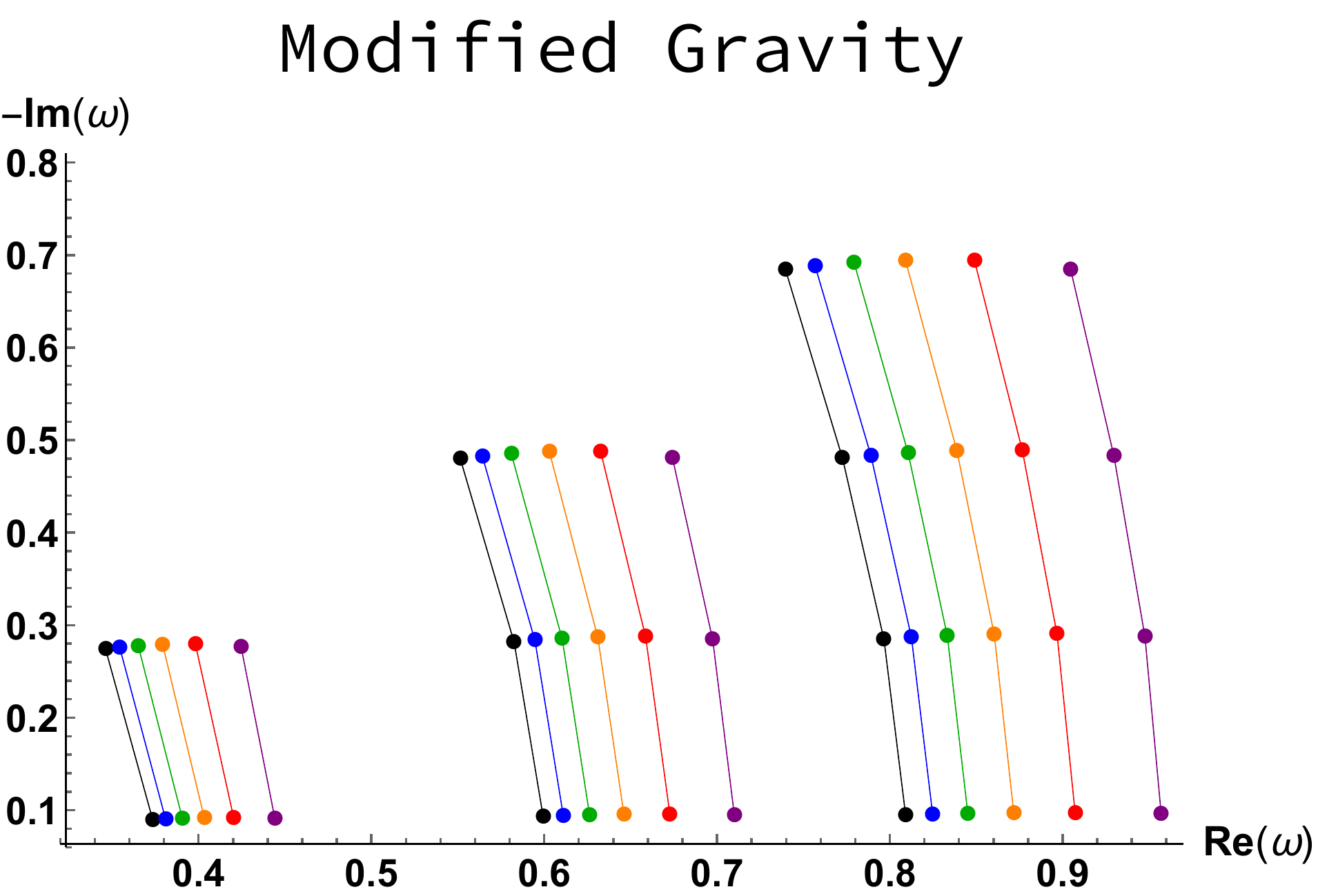}  
 \caption{QNMs in modified gravity. The left block is for $l=2$, the middle one corresponds to $l=3$ and the right one is for $l=4$.  The dark points correspond to the Schwarzschild QNMs. From left to right: $\alpha=\{1,2,3,4,5\} \times 10^{-1}$.  }
 \label{Mog}
  \end{figure}

The results are given in Fig. \ref{Mog}. The imaginary part of the QNMs is nearly the same whatever the value of $\alpha$: the modified metric has no effect on the damping rate. However, increasing $\alpha$ does increase of the real part, that is the frequency. The effect is important for values near the critical value $\alpha _c$. The slope of the Imaginary part versus the real one, at a given $l$ for different values of $n$, is nearly independent of $\alpha$. This slope is not directly observable but it shows how the structure of the QNMs changes with the overtone number. The curves remain here parallel one to the other: this means that increasing the deformation parameter does not change the frequency shift between overtones.

 \subsection{3. Ho\v{r}ava-Lifshitz gravity}
 
 Ho\v{r}ava-Lifshitz gravity bets on the fundamental nature of the quantum theory instead of relying on GR principles. It 
is a renormalizable UV-complete gravitational theory which is not Lorentz invariant in 3 + 1 dimensions \cite{Horava:2009uw}. The relativistic  time with its Lorentz invariance emerges only at large distances. Black hole solutions have been found \cite{Colgain:2009fe,Kehagias:2009is,Lu:2008js} and QNMs were studied \cite{Chen:2009gsa}.\\

Using the ansatz 
 \begin{eqnarray}
\label{ssm} ds^2 = - N^2(r)\,dt^2 + \frac{dr^2}{f(r)} + r^2
(d\theta^2 +\sin^2\theta d\phi^2)\,
\end{eqnarray}
in the action, one is led to the
Lagrangian 
\begin{eqnarray}
\label{react} \tilde{{\cal L}}_1&=&\frac{ \kappa^2\mu^2 N
}{8(1-3\lambda)\sqrt{f}}\Bigg(  \frac{\lambda-1}{2} f'^2 -
\frac{2\lambda (f-1)}{ r}f' \\
&+& \frac{(2\lambda-1)(f-1)^2}{ r^2}-2w(1
- f - r f')\Bigg),
\end{eqnarray} wherei
$w=8\mu^2(3\lambda-1)/\kappa^2$. For
$\lambda=1$, the solution is 

 \be 
 N^2=f(r)= \frac{2(r^2 -2Mr + \beta )}{r^2 +2 \beta + \sqrt{r^4 + 8 \beta M r}} ,
 \ee
 with $\beta=1/(2w)$, $w$ being the deformation parameter enterring the action given in \cite{Kehagias:2009is}. There are two horizons for $M^2 > \beta$.\\

\begin{figure}
\centering
    \includegraphics[width=.9\linewidth]{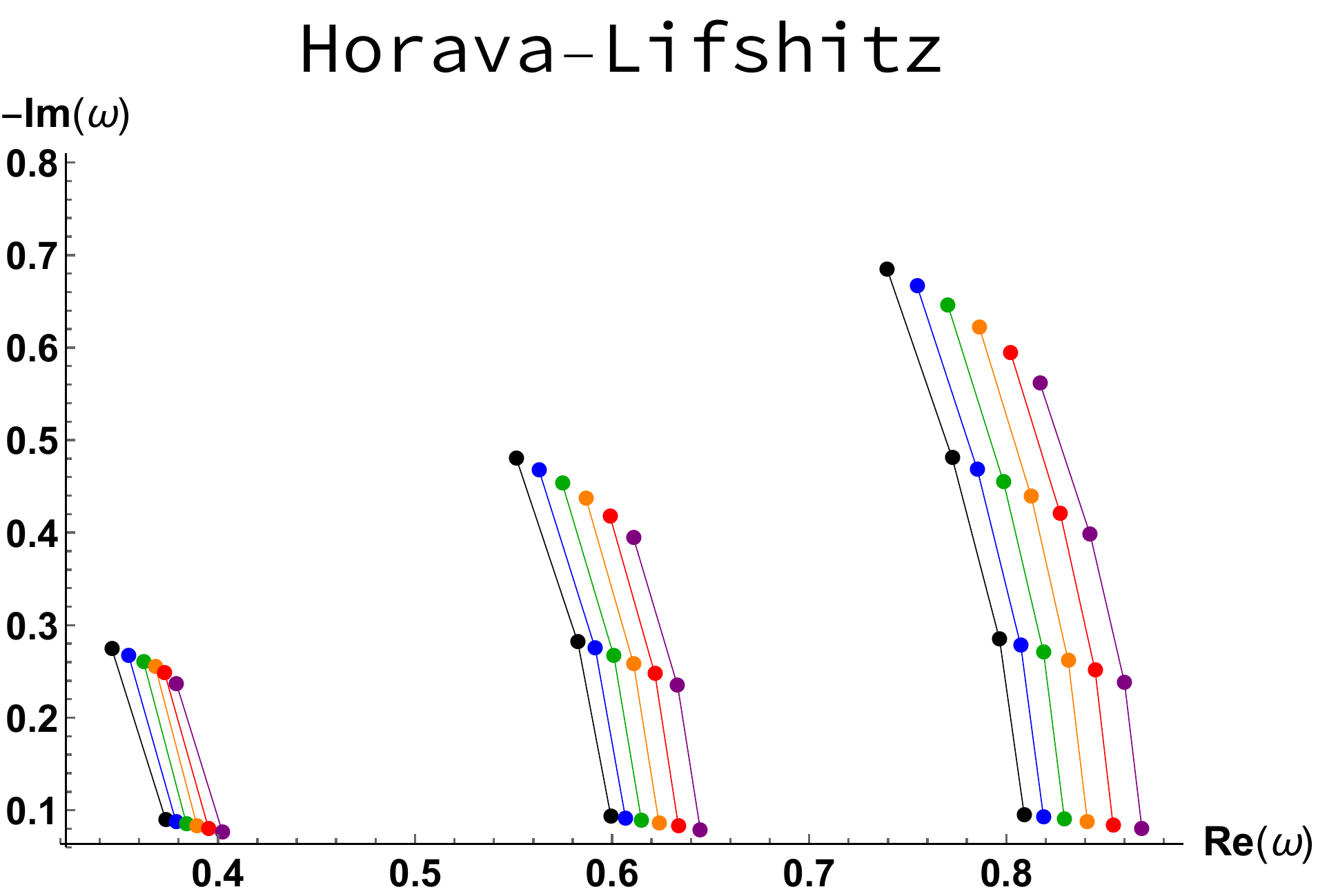}  
 \caption{QNMs in Horava-Lifshits gravity. The left block is for $l=2$, the middle one corresponds to $l=3$ and the right one is for $l=4$. The dark points correspond to the Schwarzschild QNMs. From left to right: $\beta= \{15, 30, 45, 60, 75 \} \times 10^{-2}$.}
 \label{HL}
 
 \end{figure}

The results are given in Fig. \ref{HL}. The frequency increases with an increase of $\beta$. Interestingly, the imaginary part of the overtones is highly sensitive to $\beta$. This remains true for higher multipoles. The relative variation of the imaginary part is nearly the same whatever the overtone number.  It therefore becomes large in absolute value for high $n$ values. 

\subsection{4. $\hbar$ correction}

It has been known for a long time that quantum corrections to the Newtonian gravitational potential can be rigorously derived without having a full quantum theory of gravity at disposal (see, {\it e.g.}, \cite{Donoghue:1993eb,Donoghue:1994dn,BjerrumBohr:2002ks,BjerrumBohr:2002kt,Bjerrum-Bohr:2014zsa} to cite only a few works from a very long list). Recently, a quite similar approach was developed \cite{Bargueno:2016qhu} requiring that the quantum mechanically-corrected metric reproduces the corrected Newtonian limit,  reproduces the standard result for the entropy of black holes including the known corrections, and fulfills some consistency conditions regarding the geodesic motion.\\

The resulting metric is




 \be 
 f(r)= 1-\frac{2M}{r}+\gamma \frac{2M}{r^3}.
 \ee

We use, as previously, natural units and the coefficients of the last term, $\gamma$, is proportional to $\hbar$ in these models.
It is worth noticing that there has been a long controversy about the value and the sign of the $\gamma$ factor. From the phenomenological perspective, we do not fix it to a particular value but we keep it negative, in agreement with the latest expectations.\\

\begin{figure}
\centering
    \includegraphics[width=.9\linewidth]{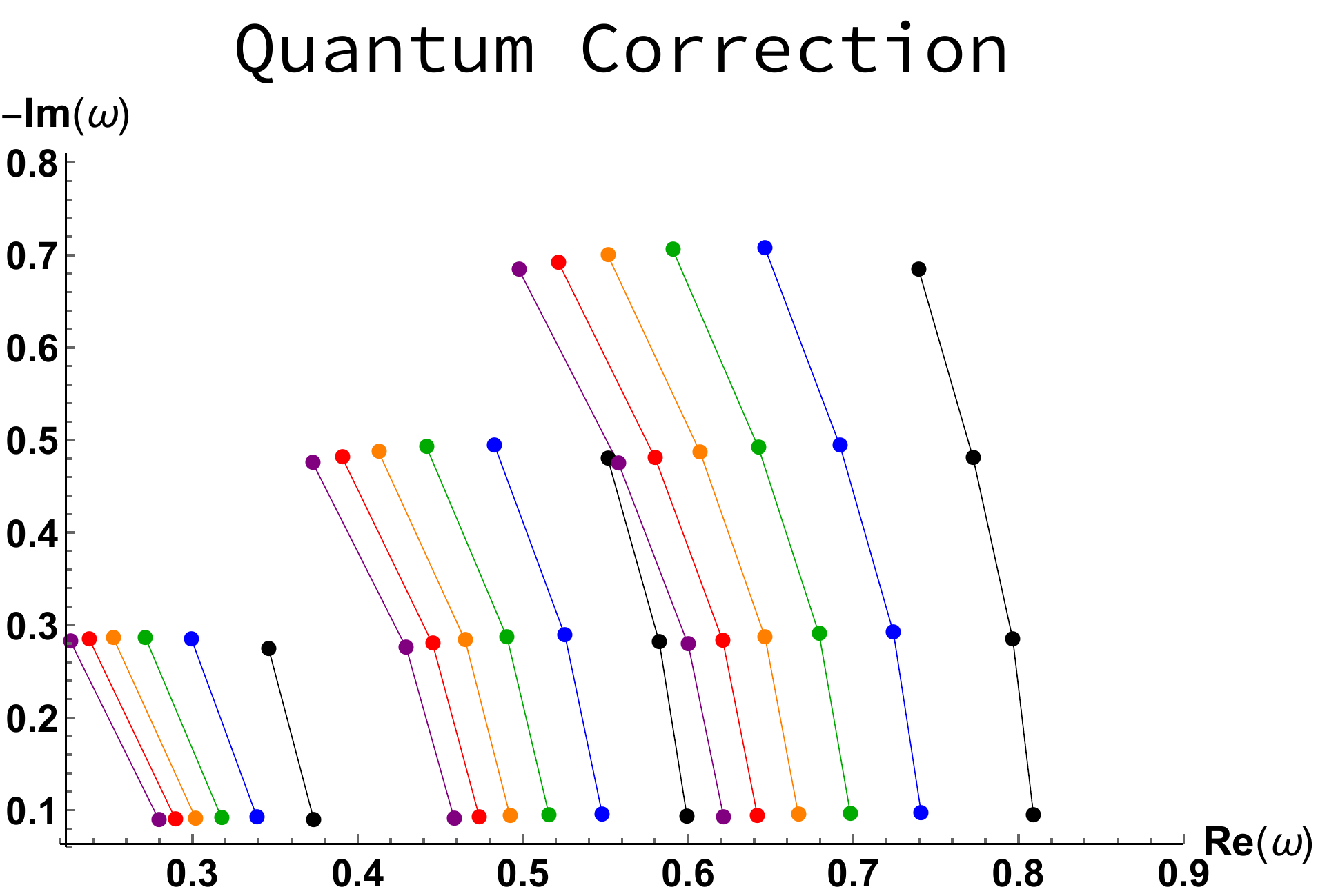}  
 \caption{QNMs in quantum-corrected gravity. The left block is for $l=2$, the middle one corresponds to $l=3$ and the right one is for $l=4$.  The dark points correspond to the Schwarzschild QNMs. From left to right: $\gamma=\{-5,-4,-3,-2,-1\}$. }
 \label{Batic}
  \end{figure}

The results are given in Fig. \ref{Batic}. For large values of $\gamma$, the effects are noticeable on the frequency. It is remarkable that, from our analysis, the real part of the complex frequency is only decreased, which is not the case for the other models that have been considered in this study. The higher the absolute value of  $\gamma$, the larger the difference of frequency between the fundamental and the overtones. This effect however remains quite subtle.
 
 \subsection{5. LQG polymeric BH}

Loop Quantum Gravity (LQG) is a non-perturbative and background-independant quantum theory of gravity \cite{Ashtekar:2012np}. In the covariant formulation, space is described by a spin network \cite{Rovelli:2011eq}. Each edge carries a ``quantum of area", labelled by an half integer $j$, associated with an irreducible representations of $SU(2)$. Each node carries a ``quantum of space" associated with an intertwiner. A key result is that area is quantized according to 

\be
A(j)= 8 \pi \gamma _{BI} \sqrt{j(j+1)},
\label{area}
\ee

\noindent with $\gamma _{BI}$ the Barbero-Immirzi parameter. Black holes are usually described in LQG through an isolated horizon puncturing a spin network \cite{Perez:2017cmj} and the phenomenology is very rich, depending on the precise setting chosen \cite{Barrau:2018rts}. We focus here on the model developed in \citep{Alesci:2012zz}, as this is the one leading to metric modifications outside the horizon, where a regular lattice with edges of lengths $\delta _b$ and $\delta _c$ is considered. Requiring the minimal area to be one derived in LQG, one is left with only one free parameter $\delta$. From this minisuperspace approximation, a static spherical solution can be derived and is given by 

\bea
&& ds^2 = - G(r) dt^2 + \frac{dr^2}{F(r)} + H(r) d\Omega^2~, \nonumber \\
&& G(r) = \frac{(r-r_+)(r-r_-)(r+ r_{*})^2}{r^4 +a_o^2}~ , \nonumber \\
&& F(r) = \frac{(r-r_+)(r-r_-) r^4}{(r+ r_{*})^2 (r^4 +a_o^2)} ~, \nonumber \\
&& H(r) = r^2 + \frac{a_o^2}{r^2}~,
\label{g}
\eea
where $d \Omega^2 = d \theta^2 + \sin^2 \theta d \phi^2$, $r_+ = 2m$ and $r_-= 2 m P^2$ are the two horizons, and $r_* = \sqrt{r_+ r_-} = 2mP$, $P$ being the polymeric function defined by $P = (\sqrt{1+\epsilon^2} -1)/(\sqrt{1+\epsilon^2} +1)$, with $\epsilon=\gamma _{BI} \delta$, and the area parameter $a_0$ is given by $a_0=A_{min}/8 \pi$, $A_{min}$ being the minimum area appearing in LQG. The parameter $m$ in the solution is related to the ADM mass $M$ by $M = m (1+P)^2$.\\

  \begin{figure}
\centering
    \includegraphics[width=.9\linewidth]{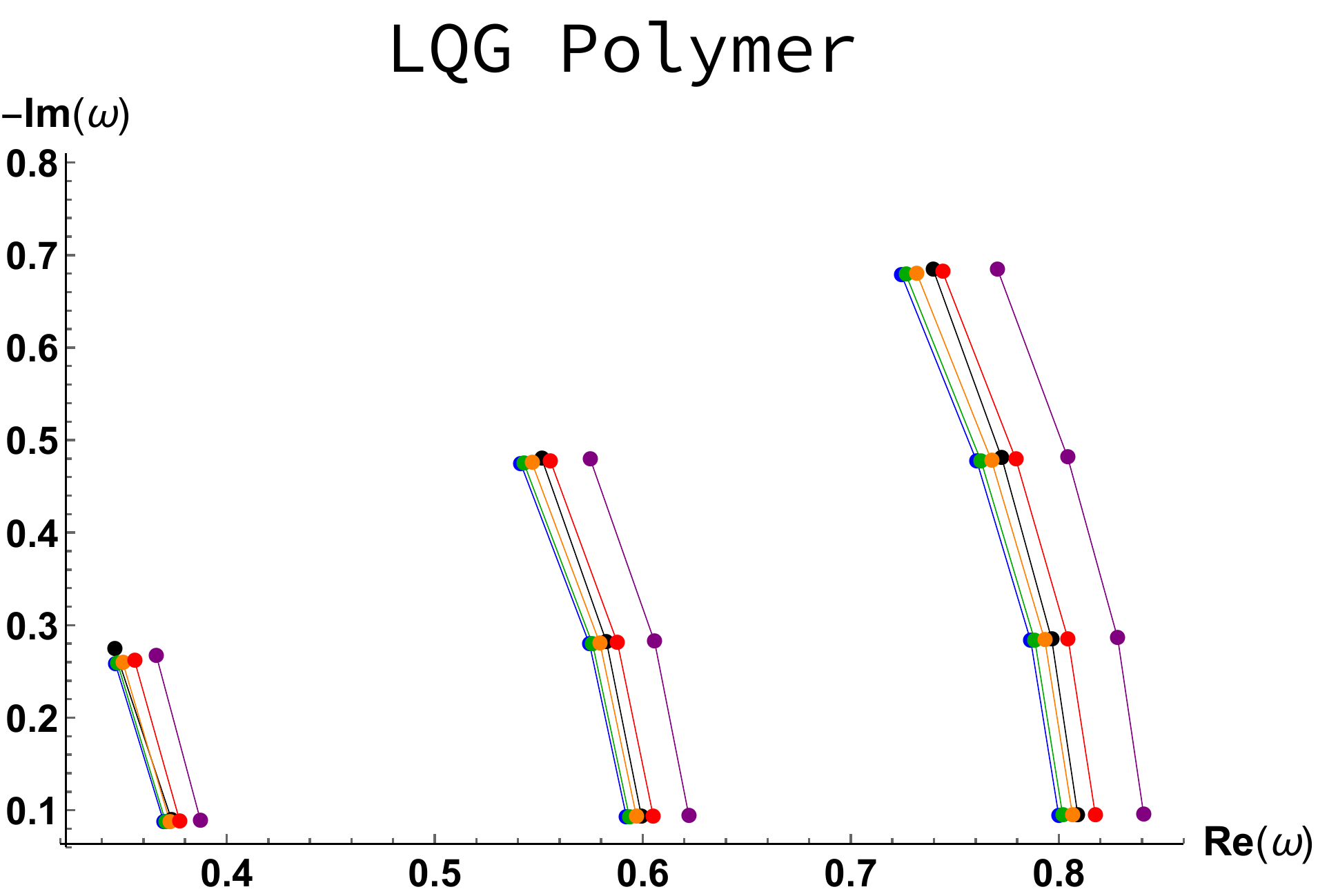}  
 \caption{QNMs in LQG (polymer BHs). The left block is for $l=2$, the middle one corresponds to $l=3$ and the right one is for $l=4$. The dark points correspond to the Schwarzschild QNMs. The parameters are $a_0=1$ and from left to right: $\epsilon=10^{-x}$ with $x\in \{-1,-0.8,-0.6,-0.4,-0.2\}$.  }
 \label{LQG}
  \end{figure}

The results are given in Fig. \ref{LQG}. The damping rate does not depend at all on the polymerization parameter. The real part of the complex frequency does, however, first decrease with $\delta$. Noticeably, the slop is unchanged and varying the deformation parameter just lead to a horizontal translation of the QNM frequency in the complex plane. This means that the frequency shift between the fundamental and the overtones does not depend on the amplitude of the quantum gravity corrections, as in modified gravity. Interestingly, for higher values of $\delta$, the frequency begins to increase. This is the only model considered in this study with a non-monotonic behavior. For $\delta\approx 10^{-0.7}$ the ``polymerization" effect nearly exactly compensates the ``area discretization" effect and one recovers the GR frequencies (and damping rates).

\section{Conclusion}

This study shows the evolution of the complex frequency of quasinormal modes of a Schwarzschild black hole for the fundamental and the first overtones for a few multipole numbers. We have considered massive gravity, STV gravity, Ho\v{r}ava-Lifshitz gravity, quantum corrected gravity, and loop quantum gravity. All the results were derived using the very same WKB approximation scheme which makes a meaningful comparison possible. It will be especially useful for future quantitative studies.\\

Obviously, distinguishing between those models with observations is more than challenging. First, because there exist degeneracies, for given overtone and multipole numbers, between the models -- when taking into account that the values of the parameters controlling the deformation are unknown. Second, because the intrinsic characteristics of the observed black holes are also unknown, which induces other degeneracies. In addition, this study should be extended to Kerr black hole, which also adds some degeneracies in addition to the complexity.\\

Some interesting trends can however be underlined. For all models, the effect of modifying the gravitational theory are more important for the real part than for the imaginary part of the complex frequency of the QNMs. Otherwise stated, the frequency shift is more important than the change in the damping rate. 
Obviously, it does not make sense to quantitatively compare the results from various models as the deformation parameters are different. However, the ``trends" are clearly specific to each studied theory and there is no need to define comparable ``steps" in the deformation parameters (which do not have the same units anyway) to draw significant conclusions about the directions in which the different models considered deviate from GR.
In addition, the sign of the frequency shift, and its dependance upon the overtone and multipole numbers is characteristic of a given extension of GR. The accurate patterns are never the same, which is an excellent point for phenomenology. It can basically be concluded that a meaningful use of  QNMs to investigate efficiently modified gravity requires the measurement of several relaxation modes. This is in principle possible \cite{Sathyaprakash:2012jk} but way beyond the sensitivity of current interferometers. If features beyond GR were to be observed, the direction of the frequency shift in the complex plane would already allow to exclude models, as this article shows. The goal of this study was not to perform a detailed analysis of the discrimination capabilities of gravitational wave experiments: it simply aimed at exhibiting the main tendencies for currently considered extended gravity models, as an introduction to this special issue on ``probing new physics with black holes".

\section{Acknowledgments}

K.M. is supported by a grant from the CFM foundation. 

\bibliography{QNM}

 \end{document}